\documentstyle[12pt]{article}
\newcommand{\sect}[1]{\setcounter{equation}{0}\section{#1}}

\newcommand{\nrm}[1]{{\protect\normalsize{#1}}}

\newcommand{\LAP}{ 
{\small E}\nrm{N}{\large S}{\Large L}{\large A}\nrm{P}{\small P}
}

\newcommand{\be}{\begin{equation}}
\newcommand{\ee}{\end{equation}}
\newcommand{\bea}{\begin{eqnarray}}
\newcommand{\ena}{\end{eqnarray}}
\newcommand{\beano}{\begin{eqnarray*}}
\newcommand{\enano}{\end{eqnarray*}}
\newcommand{\nonu}{\nonumber \\}
\newcommand{\vs}[1]{\rule[- #1 mm]{0mm}{#1 mm}}
\newcommand{\hs}[1]{\hspace{#1 mm}}
\newcommand{\A}{\alpha}

\newcommand{\lda}{\lambda}
\newcommand{\sig}{\sigma}

\newcommand{\eps}{\epsilon}

\newcommand{\vph}{\varphi}
\newcommand{\bal}{{\overline{\alpha}}}
\newcommand{\bbet}{{\overline{\beta}}}

\newcommand{\cf}{\mbox{${\cal F}$}}
\newcommand{{\cg}}{\mbox{$\cal{G}$}}
\newcommand{{\tcg}}{\mbox{$\wt{\cal{G}}$}}
\newcommand{{\bcg}}{\mbox{$\soul{\cal{G}}$}}

\newcommand{{\tch}}{\mbox{$\wt{\cal{G}}_0$}}

\newcommand{\cp}{\mbox{$\cal{P}$}}

\newcommand{\cs}{\mbox{$\cal{S}$}}

\newcommand{\cu}{\mbox{${\cal U}$}}
\newcommand{\cw}{\mbox{$\cal{W}$}}

\newcommand{\su}{\mbox{${s\ell(2)}$}}

\newcommand{\prt}{\partial}

\newcommand{\soul}[1]{\overline{#1}}

\newcommand{\wt}[1]{\widetilde{#1}}
\newcommand{\mb}[1]{\hs{4}\mbox{#1}\hs{4}}
\newcommand{\half}{\frac{1}{2}}

\newcommand{\pcr}{\mbox{Poincar\'e }}

\newcommand{\ie}{{\it i.e.}\ }
\newcommand{\C}{\mbox{\hspace{.0em}C\hspace{-.842em} I\hspace{.24em}}}
\newcommand{\R}{\mbox{\hspace{.0em}I\hspace{-.42em} R}}
\newcommand{\Q}{\mbox{\hspace{.0em}Q\hspace{-.942em} I\hspace{.3em}}}
\newcommand{\Z}{\mbox{$Z$\hspace{-0.42em}$Z$}}

\newcommand{\1}{\mbox{\hspace{.0em}1\hspace{-.24em}I}}
\newcommand{\NP}[1]{Nucl.\ Phys.\ {\bf #1}}

\newcommand{\CMP}[1]{Comm.\ Math.\ Phys.\ {\bf #1}}
\newcommand{\JMP}[1]{Journ.\ Math.\ Phys.\ {\bf #1}}
\newcommand{\IJMP}[1]{Int. Journ.\ Mod.\ Phys.\ {\bf #1}}

\begin{document}
\renewcommand{\thefootnote}{\fnsymbol{footnote}}
\newpage
\pagestyle{empty}
\setcounter{page}{0}

\null
\begin{minipage}{4.9cm}
\begin{center}
{\bf  G{\sc\bf roupe} d'A{\sc\bf nnecy}\\ \ \\
Laboratoire d'Annecy-le-Vieux de Physique des Particules}
\end{center}
\end{minipage}
\hfill
\hfill
\begin{minipage}{4.2cm}
\begin{center}
{\bf G{\sc\bf roupe} de L{\sc\bf yon}\\ \ \\
Ecole Normale Sup\'erieure de Lyon}
\end{center}
\end{minipage}

\begin{center}
\rule{14cm}{.42mm}
\end{center}

\vfill

\begin{center}

{\LARGE {\bf \cw-realization of Lie algebras:\\[.42cm]
Application to $so(4,2)$ and \pcr algebras}}\\[1cm]

\vs{2}

{\large F. Barbarin$^{1}$, E. Ragoucy$^{1}$, 
and P. Sorba$^{1,2}$}

{\em Laboratoire de Physique Th\'eorique }\LAP\footnote{URA 
14-36 du CNRS, associ\'ee \`a l'Ecole Normale Sup\'erieure de 
Lyon  et \`a l'Universit\'e de Savoie, 

\noindent
$^1$ Groupe d'Annecy: LAPP, Chemin de Bellevue BP 110, 
F-74941 Annecy-le-Vieux Cedex, France.

\noindent
$^2$ Groupe de Lyon: ENS Lyon, 46 all\'ee d'Italie, F-69364 
Lyon Cedex 07,France.
}\\
\end{center}
\vfill

\centerline{ {\bf Abstract}}

\indent
The property of some finite \cw-algebras to appear as the commutant of a
particular subalgebra in a simple Lie algebra \cg\ is exploited for the
obtention of new \cg-realizations from a "canonical" differential one.

The method is applied to the conformal algebra $so(4,2)$ and therefore
yields also results for its Poincar\'e subalgebra. Unitary irreducible
representations of these algebras are recognized in this approach, which
is naturally compared -or associated- to the induced representation
technic.
\vfill
\centerline{\it{ Submitted to \CMP{}}}
\vfill
\rightline{hep-th/9606014}
\rightline{\LAP-AL-594/96}
\rightline{May 1996}

\newpage
\pagestyle{plain}
\setcounter{page}{1}
\setcounter{footnote}{0}
\renewcommand{\thefootnote}{\arabic{footnote}}
\sect{Introduction}
As shown in \cite{BRS,nonPol}, the construction of finite
\cw-algebras achieved in the framework of Hamiltonian reduction \cite{BT}
also leads to the determination of the commutants, in the enveloping
algebra\footnote{Strictly speaking, we consider a generalization
(localization) of \cu(\cg), which contains, apart from \cu(\cg) itself,
quotients $u^{-1}v$, $vu^{-1}$, and also $u^r$, $r\in\Q$, where
$u,v\in\cu(\cg)$, $u\neq0$.}
 $\cu(\cg)$, of particular subalgebras of a simple Lie algebra \cg. 
 
Such a property can be used to build, from a special
realization of \cg, a large class of \cg-representations. Let us be more
specific, and consider a graded decomposition of the simple Lie algebra
\cg\ of the type
\be
\cg=\cg_{-1}\oplus \cg_{0}\oplus \cg_{+1}
\ee
with the corresponding commutation relations:
\be
{[\cg_{0},\cg_{\pm1}]}\subset\cg_{\pm1} \mb{;} 
{[\cg_{\pm1},\cg_{\pm1}]}=\{0\} \mb{;} 
{[\cg_{+1},\cg_{-1}]}\subset\cg_{0} 
\ee
As reminded in section \ref{Wcom}, it is possible to recognize in
the commutant of a \cg-subalgebra \tcg\ of the form $\tcg=\cg_{-1}\oplus
\tcg_{0}$, with $\tcg_{0}\subset\cg_{0}$, a particular finite
\cw-algebra. The determination of the realization of the \cw-algebra needed 
for our purpose can be obtained in a systematic way after some modification
\cite{nonPol} of the usual Hamiltonian reduction technic \cite{ORaf}. 
Moreover, one knows how to construct a
realization of the \cg\ with differential operators on the space of
smooth functions $\vph(y_1,\dots,y_s)$ with $s=dim\cg_-$. In this
picture, the abelianity of the $\cg_{-1}$-part allows  
each $\cg_{-1}$ generator to act by direct multiplication:
\be
\vph(y_1,\dots,y_s) \ \rightarrow\ y_i\vph(y_1,\dots,y_s)
\mb{with} i=1,\dots,s
\ee
-cf action of the translation group- while the generators of the
$\cg_0\oplus\cg_+$ part will be represented by polynomials in the $y_i$
and $\prt_{y_i}$ (see section \ref{Wreal}).

It is from this particular -canonical- differential realization of \cg\
that new representations will be constructed with the use of the finite
\cw-algebra above mentioned. Realization of the \tcg\ generators will not
be affected in this approach. On the contrary, to the differential form
of each generator in the $\cg\setminus\tcg$ part will be added a sum of
\cw-generators, the coefficients of which are functions
$f(y_i,\prt_{y_i})$. By associating a matrix differential realization to
each irreducible finite dimensional \cw-representation, one will get an
action of \cg\ on vector functions $\vec{\vph}=(\vph_1,\dots,\vph_d)$,
with $\vph_i=\vph_i(y_1,\dots,y_s)$, where $d$ is the dimension of the
considered \cw-representation.

Such an approach has already been considered with some success \cite{BRS}
for the study of the Heisenberg quantization for a system of two
particles in 1 and 2 dimensions \cite{LM}. The corresponding algebras are
respectively $sp(2)$ and $sp(4)$, and in each case, it has been possible
to relate the anyonic parameter to the eigenvalues of a \cw-generator.

It has seemed to us of some interest to apply our technic on a rather
well-known algebra, the $so(4,2)$ one, also called the 4-dimensional
conformal algebra. The decomposition of interest is the one in which the
$\cg_{-1}$ part is constituted by the four translations, $\cg_0$ by the
Lorentz algebra $so(3,1)$ plus the dilatation, and $\cg_+$ by the four
special conformal generators. The representation space is then the set of
smooth functions $\cf(M^4,\C)$ where $M^4$ is the usual Minkowski space,
and the finite \cw-algebra of interest appears as a non-linear
deformation of the $\cg_0$ subalgebra. This algebra is the commutant in
\cu(\cg) of the part $\cg_-\oplus\tcg_0$, where $\tcg_0$ is a four
dimensional subalgebra of $\cg_0$. Note also that the commutant of
$\cg_-$ can be seen as the direct sum of this \cw-algebra and of $\cg_-$, \ie
$\cw\oplus\cg_-$. As presented in section \ref{sect3}, a suitable basis
for this non-linear algebra is the one made of three generators, components of
the $\vec{J}$ vector constituting an $so(3)$ part, and three other
generators forming a second vector $\vec{S}$ under the $\vec{J}$ part,
the commutation relation among the $S_i$'s producing a seventh generator
belonging to the center of \cw.

It is interesting to remark that, restricting to the {\pcr} subalgebra the
general realization so obtained, only the $\vec{J}$ part contributes and
appears to be the spin algebra directly connected to the $(W_\mu)$
Pauli-Lubanski-Wigner four vector. The usual \pcr representations can
therefore easily be recognized in this approach, the \cw-part being
related to the usual Wigner rotations. Then, realizations of the special
conformal generators involve naturally the other \cw-generators. In a
rather elegant way, a $(\Sigma_\mu)$ four vector, conformal analogous to
the $(W_\mu)$ one, can be exhibited and connected to the $\vec{S}$ part
of the \cw-algebra. 

Apart from this geometrical picture, one can recognize through the finite
dimensional representation of the \cw-algebra, the characterization of
the unitary irreducible representations of the conformal algebra, as
computed in \cite{Mack} in case of positive energy. 

It is clear, from the considered example, that the proposed method
provides the same results as those given by the usual induced
representation technics. This is discussed at the end of the paper.

\sect{\cw-algebras as commutants\label{Wcom}}
\subsection{General results}
We recall here the framework which leads to
the main result obtained in \cite{nonPol}, namely

\indent

{\bf{Theorem}}

{\em Any finite \cw(\cg,\cs)\ algebra, with 
$\cs=\mu\su$ regular subalgebra of \cg,
can be seen as the
commutant in the enveloping algebra
$\cu(\cg)$ of some \cg-subalgebra $\wt{\cg}$. 

Moreover, let $H$ be the Cartan generator of the diagonal \su\
 in \cs. If we call $\cg_-$, $\cg_0$ and $\cg_+$ the
eigenspaces of respectively negative, null, and positive eigenvalues
under $H$, then
\tcg\ decomposes as $\tcg=\cg_-\oplus\tch$, where $\tch$ is a
subalgebra of $\cg_0$ which can be uniquely determined (see
below).}

\indent

We start with a simple Lie algebra \cg, and consider an \su-embedding
defined through the regular subalgebra \cs\ in \cg. Together with this
\su-embedding, we get a gradation of \cg, using the action of the \su\
Cartan generator $H$. We will call $\cg_-$, $\cg_0$ and $\cg_+$ the
eigenspaces of respectively negative, null, and positive eigenvalues
under $H$. Below, we will consider the case where 
\cs\ is a (sum of) regular subalgebra(s) \su,
which leads to Abelian $\cg_-$ subalgebras. 

In \cite{nonPol}, we have shown that, for integral gradations,
the commutant of $\cg_-$ in
the enveloping algebra $\cu(\cg)$ is isomorphic to the 
$\cw(\cg,\mu\,\su)$ algebra plus a center. 
Moreover, we have also shown that one can get rid of the center when
looking at the commutant of a wider subalgebra
$\tcg=\cg_-\oplus\tch$ where $\tch$ is a subalgebra of $\cg_0$
and $\cg_-=\cg_{-1}$. In the case of half-integral gradations, we should use
the halving procedure. However, when $\cs=\mu\su$, it can be shown that we 
still have $\tcg=\cg_-\oplus\tch$ with $\cg_-=\cg_{-1} \oplus \cg_{-\half}$.

Note that more general finite \cw(\cg,\cs) algebras may be obtained as a
commutant, but one has to extend \tch\ to the $\cg_+$ part of \cg\ (see
example in \cite{nonPol}). The case of "affine" \cw(\cg, \cs) algebra can
also be treated with this framework \cite{nonPol}.

As an illustration, let us consider $\cs=\mu\,\su$ in
$\cg=s\ell(n)$. The simple roots of \cg\ will be noted
$\alpha_i$ with $1\leq i\leq n-1$. Then,
one can take $\tcg=\oplus_{i=1}^\mu \Gamma_i$, where $\Gamma_i$
is the subalgebra 
\be
\Gamma_i=\{E_{\beta_j},\ i\leq j\leq n-i\ ;\ E_{\gamma_j},\ 
i\leq j\leq n-i-1\ ;\ H_{\beta_{n-i}}\} \mb{with}
\left\{\begin{array}{l}\beta_j=\sum_{k=i}^j \alpha_k\\
\gamma_j=\sum_{k=i}^j \alpha_{n-k}\end{array}\right. \label{gtild}
\ee

\subsection{Explicit calculations}
We start with a general element $J=J^at_a$ in \cg.  $J^a$ are the
coordinate-functions ($J^a\in\cg^*)$, and it is known that one can
provide them  with a Poisson Bracket structure that mimics the
commutation relations of \cg: $[t_a,t_b]={f_{ab}}^ct_c$ while
$\{J^a,J^b\}={f^{ab}}_c J^c$, indices being lowered via the
Killing metric $\eta_{ab}=<t_a,t_b>$, 
and raised by its inverse $\eta^{ab}$. Thanks to this metric (and these
PB) we identify $\cg^*$ with \cg\ so that the decompositions of \cg\ 
(like $\cg=\cg_-\oplus\cg_0\oplus\cg_+$) 
will be also valid for $\cg^*$, that
we will loosely write \cg. We will also use the Cartan involution
$\Theta$ defined by $\Theta(E_{\alpha_i})=-E_{-\alpha_i}$ for any simple
root $\alpha_i$. It allows us to define another scalar product
$(X,Y)=<X,\Theta(Y)>$.

We consider the decomposition
$\cg=\cg_-\oplus\cg_0\oplus\cg_+$ obtained by the Cartan generator of the
diagonal $\su_{diag}$ in $\mu$ regular \su.
As a notation, we use latin indices $a,b,c,\dots$ to label the
generators of \cg, and greek (resp. overlined-greek) to label generators
in  \tcg\ 
(resp. in $\soul{\cg}$, its orthogonal complementary w.r.t. the scalar
product $(.,.)$ ). When needed, we will also use $t_A$, $t_i$ and
$t_{\soul{A}}$ to denote the generators of $\cg_-$, $\cg_0$, and
$\cg_+$ respectively.
Thus, we note 
\be
J=J^at_a=J^\alpha t_\alpha+J^\bal t_\bal=J^A t_A+J^i t_i+
J^{\soul{A}} t_{\soul{A}} \mb{with}\left\{
\begin{array}{l} t_a\in\cg \\ 
t_\alpha\in\tcg \mb{;} t_\bal\in\bcg \\
t_A\in\cg_- \mbox{ ; } t_i\in\cg_0 \mbox{ ; } t_{\soul{A}}\in\cg_+ 
\end{array}\right.
\ee

To compute the commutant of \tcg\ in \cu(\cg), we consider the group
transformations $J\rightarrow J^g=gJg^{-1}$ with 
$g\in \Theta(\wt{G})$
where $Lie\Theta(\wt{G})=\Theta(\tcg)$, $\wt{G}$ being the connected
component of the Lie group whose algebra is \tcg.  
In fact, if $g=e^{x^b t_b}$, with $t_b\in\Theta(\tcg)$, one can
rewrite $gJg^{-1}$ as $exp\left(x_b\left\{ J^b,.\right\}\right)(J)$
where now the PB  $\left\{J^b,.\right\}$ acts on $J^a$ (instead of
$t_a$),  $J^b$ belongs to $\tcg^*$, and
$x_b=\eta_{bc}x^c$. Then, one can show that, starting
from any generic element $J$, there exists a unique group generator 
$g=e^{x^b t_b}\in \Theta(\wt{G})$ 
such that $J^g=e_-+W^{\bal} t_{\bal}$, where $e_-$ is the negative root
generator of the $\su_{diag}$ under consideration and
$t_{\bal}\in\bcg$. Let us remark that the
parameters $x^b$ depend only on the components $J^\A\in\tcg$.

In fact, decomposing $\Theta(\wt{G})$ as 
$\Theta(G_-)\,\Theta(\wt{G}_0)$ (in the neighborhood of the identity), it
is easy to show that $\Theta(\wt{G}_0)$ is defined as the subgroup of $G_0$
which transforms $J$ into $J'$ with 
\be
h\in\Theta(\wt{G}_0)\ :\ J\rightarrow J^h=J'\mb{with}
\left.J'\right|_{\cg_-}=e_- \label{dur}
\ee
 while $\Theta(G_-)$ transforms
$J'$ into $J^g=e_-+W^{\bal} t_{\bal}$:
\be
g\in \Theta(\wt{G}_-)\ :\ J'\rightarrow J^g=e_- +W^{\bal} t_{\bal}
\label{dur2}
\ee
 This property is still true when $\cg_-$ is
not Abelian, but in that case one has to include elements of 
$\Theta(G_+)$ in order for $\Theta(\wt{G})$ to satisfy (\ref{dur}).

Altogether, the generators $W^{\bal}$ are the "gauge-fixed" components of
$J^g$. Hence, they have
vanishing PB with any element in \tcg, 
and generate the commutant of \tcg. They
form a classical (PB) version of the \cw-algebra.  Note that the $W^{\bal}$
are linear in $J^\bbet$: we will come back on this property in section
\ref{sect2:3}.

A quantum version of this algebra will be obtained by
using the map $i:\ \cg^*\rightarrow\cg$ defined by $i(J^a)=\eta^{ab}t_b$
and extended to \cu(\cg) by symmetrization of the products (see \cite{nonPol} 
for details).

Explicitly, one uses a matrix representation for the $t_a$ to get a
simple expression for the group generators $g\in\Theta(\wt{G})$ which act on 
$J$. We determine $g$ such that $J^g$ takes the form (\ref{dur2}). The
coefficients $W^{\bal}$ of $t_{\bal}$ are then expressed in terms of 
{\em all} the
$J^a$'s, contrarily to the usual Hamiltonian reduction approach, since we
have not imposed any constraint on the current components. Finally, from
the  classical expression of the \cw-generators, we use the above mentioned
quantization procedure to obtain the commutant of \tcg\ in \cu(\cg).

\subsection{Example: \su\label{Sectsu2}}
We explicit here the results just presented in the case of $\cg=\su$. 
We start with a generic element
\be
J=J^at_a=\left(\begin{array}{cc} \frac{1}{2}J^0 & J^+ \\
J^- & -\frac{1}{2}J^0 \end{array}\right)
\ee
As normalization, we take
\be
\begin{array}{l}
{[t_0,t_\pm]}=\pm t_\pm \mb{;} [t_+,t_-]=2t_0 \\
\eta_{00}=\half \mb{;} \eta_{+-}=1 \mb{so that} 
\eta^{00}=2 \mb{;} \eta^{+-}=1 \\
\{J^0,J^\pm\}=\pm2 J^\pm \mb{;} \{J^+,J^-\}=J^0
\end{array}
\ee
Here the eigenspaces $\cg_{\pm,0}$ are one-dimensional. 
Thus, \tch\ must be $\cg_0$
itself, and indeed we have:
\[
h=\left(\begin{array}{cc}\sqrt{J^-} & 0 \\ 
0 & \frac{1}{\sqrt{J^-}}
\end{array}\right)
\mb{is such that} hJh^{-1}=J'= \left(\begin{array}{cc} J'^0 & J'^+ \\
1 & -J'^0 \end{array}\right) \mb{with} \left\{\begin{array}{l}
J'^0=\frac{1}{2}J^0 \\ J'^+=J^+ J^- \end{array}\right.
\]
Now, acting with $G_+$ leads to
\[
gJ'g^{-1}= \left(\begin{array}{cc} 0 & W \\
1 & 0 \end{array}\right) \mb{with} W=J'^+ +(J'^0)^2={J^+}{J^-}
+\frac{1}{4}(J^0)^2
\mb{and} g=\left(\begin{array}{cc}1 & -J'^0 \\ 0 & 1\end{array}\right) 
\]
After quantization,
we recover the well-known result that the Casimir operator 
\be
C_2=t_0^2+\frac{1}{2}(t_+t_-+t_-t_+)=t_0^2+t_0+t_-t_+
\label{eq:1.5}
\ee
commutes will $t_-$ and $t_0$. Note however that we get also the
non-trivial information that the only elements in $\cu(\su)$ that commute
with both $t_-$ and $t_0$ (not necessarily $t_+$) are polynomials in
$W$.

\sect{\cw-realization of Lie algebras\label{Wreal}}
The above construction allows us to provide a new technic for the
construction of representations of a Lie algebra \cg\ starting from a given
(canonical) realization of \cg.
\subsection{Regular representation of a Lie algebra \cg}
We recall here the regular representation of a Lie algebra \cg. We start
with $G$, the Lie group of \cg, and consider $\cf=\cf(G,\C)$ the space
of smooth functions on G. One can construct two isomorphic representations of
$G$ on \cf. They are associated to the right and left actions of $G$ onto
itself.

The left representation $U_L$ is defined through the map 
\be
g\in G \rightarrow U_L(g) \mb{with}
(U_L(g)\vph)\, (g')=\vph(g^{-1}g')\ ,\forall \vph\in\cf,\ \forall g'\in G
\ee
while the right representation $U_R$ uses the map
\be
g\rightarrow U_R(g) \mb{with}
(U_R(g)\vph)\, (g') =\vph(g'g)\ ,\forall \vph\in\cf,\ \forall g'\in G
\ee
These representations induce naturally two representations of \cg\ on
\cf:
\beano
t\in\cg\rightarrow T_L(t) \mb{with}
T_L(t)\vph(g') &=& \left[\frac{d}{d\eps}
\vph(e^{-\eps t}g')\right]_{\eps=0} \\ 
t\in\cg\rightarrow T_R(t) \mb{with}
T_R(t)\vph(g') &=& 
\left[\frac{d}{d\eps} \vph(g'e^{\eps t})\right]_{\eps=0}  
\enano
From now on, we will focus on $U_R$ and $T_R$, the way the following
construction can be done for $U_L$ and $T_L$ being analogous.

\subsection{Differential realization of \cg\label{diffReal}}
We now use the regular representation to build a differential realization
(as used in \cite{Ko}) adapted to the decomposition\footnote{In the case
of half-integral gradation $\cg_-=\cg_{-1}\oplus\cg_{-\half}$ we will 
perform a halving to
get an integral one.}
$\cg=\cg_-\oplus\cg_0\oplus\cg_+$
that we have used in the previous section. We first remark that
$\cp=\cg_0\oplus\cg_+$ is a parabolic subalgebra of \cg, and that any
element $g\in G$ can be decomposed in a unique way as $g=pg_-$ with
$p\in P$ and $g_-\in G_-$.

Now, starting from the regular representation $T_R$ on \cf, we can
restrict it to the subspace $\soul{\cf}$ of functions $\soul{\vph}\in\cf$
such that $\soul{\vph}(pg)=\soul{\vph}(g)$ $\forall g\in G$ and
$\forall p\in P$. We will call
this realization $\soul{T}_R$. Note that $\soul{\cf}$ is not empty because 
$\cf(\cg_-,\C)$ can be canonically embedded in $\soul{\cf}$.
Thus $\soul{T}_R$ is not the trivial representation.

Any $g\in G$ can be decomposed into $g=pg_-$ with $p\in P$ and $g_-\in
G_-$, in a unique way. Moreover, the set
$\{t_A\}$ being a basis of $\cg_-$, we can label any $g_-\in G_-$ with
$dim\cg_-=s$ variables $x^A$ such that $g_-=e^{x^A t_A}$.
Thus we will have
\be
\soul{\vph}(g)=\soul{\vph}(pg_-)= 
\soul{\vph}(g_-)=\soul{\vph}(e^{x^A t_A})
\equiv\soul{\vph}(x^1,x^2,\dots,x^s)
\ee

Let us look more carefully at $\soul{T}_R$. For $t_B\in\cg_-$, we
have: 
\be
\soul{T}_R(t_B)\soul{\vph}(g)=\left[\frac{d}{d\eps}
\soul{\vph}(ge^{\eps t_B})\right]_{\eps=0}= \left[\frac{d}{d\eps} 
\soul{\vph}(pg_-e^{\eps t_B})\right]_{\eps=0}=
\left[\frac{d}{d\eps} \soul{\vph}(g_-e^{\eps t_B})\right]_{\eps=0}
\ee
where we have used the decomposition $g=pg_-$.  Now, as $\cg_-$
is Abelian, we have
\[
\soul{\vph}(ge^{\eps t_B})=\soul{\vph}(e^{x^A t_A+\eps
t_B})=\soul{\vph}(x^1,x^2,\dots,x^B+\eps,\dots,x^s)
\]
so that
\be
\begin{array}{l}
\soul{T}_R(t_B)\vph(x^1,\dots,x^s)=\left[\frac{d}{d\eps}
\soul{\vph}(x^1,x^2,\dots,x^B+\eps,\dots,x^s)\right]_{\eps=0}=
\frac{\prt}{\prt x^B}\soul{\vph}(x^1,\dots,x^s) \\
\Rightarrow\ \soul{T}_R(t_B)=\prt_B
\end{array}
\ee
To be complete, let us add that, using
\be
e^{-B}e^Ae^B=e^{e^{-B}Ae^{B}}
\ee
the same calculation shows that
\be
{\forall} t_j\in\cg_0,\ \soul{T}_R(t_j)=-{f_{jA}}^B
x^A\prt_B \mb{where} {[t_j,t_A]}={f_{jA}}^B t_B
\ee
Then, it is easy to check that 
\be
\soul{T}_R(t_{\soul{A}})=\frac{1}{2}{f_{\soul{A}A}}^j{f_{jB}}^C
x^Ax^B\prt_C 
\ee
is solution of 
${[t_j,t_{\soul{A}}]}={f_{j{\soul{A}}}}^{\soul{B}} t_{\soul{B}}$ and
${[t_A,t_{\soul{A}}]}={f_{A{\soul{A}}}}^i t_i$.

We have seen that one can identify $\soul{\cf}$ with
 a space of functions  of $dim\cg_-$ variables. 
 We would like that $\cg_-$ elements act multiplicatively instead of acting
 through differential operators.
Thus, we perform a Fourier transformation:
\beano
\soul{\vph}\in\soul{\cf} &\rightarrow& \wt{\vph}\in\wt{\cf}\\
\soul{\vph}(x^1,\dots,x^s) &\rightarrow& \wt{\vph}(y_1,\dots,y_s)=\int
d^sx\ e^{i\vec{y}\cdot\vec{x}}\ \soul{\vph}(x^1,\dots,x^s)
\enano
with $\vec{y}\cdot\vec{x}=y_A x^A$.
On the new space of functions $\wt{\cf}$, the $\cg_-$ part acts
multiplicatively, and the other generators of \cg\ will be differential 
polynomials in the variables:
\beano
t_A\in\cg_- &\rightarrow& \wt{T}_R(t_A)=-i y_A \\
t_j\in\cg_0 &\rightarrow& \wt{T}_R(t_j)=
{f_{jA}}^B y_B\prt^A +{f_{jA}}^A 
\mb{with} \prt^B=\frac{d}{dy_B}\\ 
t_{\soul{A}}\in\cg_+ &\rightarrow& 
\wt{T}_R(t_{\soul{A}})=\frac{i}{2}{f_{\soul{A}A}}^j{f_{jB}}^C
y_C\prt^A\prt^B  +\frac{i}{2}({f_{\soul{A}A}}^j{f_{jB}}^B+
{f_{\soul{A}B}}^j{f_{jA}}^B)\prt^A
\enano
It is this representation we will use in the following.

\subsection{\cw-realization\label{sect2:3}}
We use the technics of the section \ref{Wcom} to get a wide class of
representations of \cg, starting from the differential representation
based on $\cg_-$. We know that the generators of 
$\cw(\cg,\mu\su)\equiv\cw$ form
the commutant of $\tcg=\cg_-\oplus\tch$. They can be expressed in terms of all
the generators of $\cg$.
Moreover, we have seen that, in
the differential realization, the elements of $\cg_-$ are represented by
$y_A$, while the generators in \tch\ are represented by
$y_B\prt^A$ (up to some constants). 
Thus, the \cw-generators commute both with $y_A$
and $\prt^A$ $(\forall\ A$), \ie they are constant on the whole
differential representation. Let $w_0^{\bal}$ be these constants, 
where ${\bal}$ labels the generators of \cw.

From the general results of section \ref{Wcom}, we know that there exists a
unique element $g\in\Theta(\wt{G})$ such that 
\be
J^g=g(J^\A t_\A+J^\bal t_\bal)g^{-1}=e_-+W^{\bal}t_{\bal}
\mb{with} g=e^{u^a t_a},\ t_a\in\Theta(\tcg)\ \mbox{ and }
u^a=u^a(J^\A)  \label{W(J)}
\ee
 Let us rewrite this formula as
\be
J=J^\A t_\A+J^\bal t_\bal=g^{-1}(e_-+W^{\bal}t_{\bal})g
 \label{J(W)}
\ee
This shows that we can write any element $J^a$ in terms
of the $J^\alpha$'s and the $W^{\bal}$'s. 
We also remark that, in the same way the realization of $W^{\bal}$ is
linear in $J^\bal$ because of (\ref{W(J)}), the $J^\bal$'s also depend linearly
in the $W^{\bal}$'s because of (\ref{J(W)}).
Then starting with a given realization of \tcg\, we construct a realization of
the whole algebra $\cg$ leaving \tcg\ unchanged and expressing the other 
generators of $\bcg$ in terms of those of \tcg\ and the \cw-algebra, by 
applying formula (\ref{J(W)}).

Performing the same calculation at the quantum level, we deduce that
starting with a given representation of $\cg$ and so of \tcg\, one can obtain
other representations of \cg\ using representations of the \cw-algebra.

It is natural to
start with the differential representation of \cg. Then, one generalizes
it into a "matrix differential representation" using a matrix
representation of \cw. More precisely, we take as representation space 
$d$ copies of
$\wt{\cf}$, where $d$ is the dimension of the \cw-representation one
chooses. Then, \cg\ will be represented by
\beano
t_A\in\cg_- &\rightarrow& \vec{T}_R(t_A)=-i y_A.\, \1\\
t_j\in\cg_0 \left\{\begin{array}{l} t_\alpha\in \tcg_0 \\ 
t_\bal\in (\cg_0\setminus\tcg_0)\end{array}\right.
 &\begin{array}{l} \rightarrow \\ \rightarrow \end{array}& 
 \begin{array}{l} \vec{T}_R(t_\alpha)=
\left({f_{\alpha A}}^B y_B\prt^A +{f_{\alpha A}}^A\right).\, \1 \\
\vec{T}_R(t_\bal)=
\left({f_{\bal A}}^B y_B\prt^A +{f_{iA}}^A\right).\, \1
+R_{\bal \bbet}(y_A,\prt^B).(M^{\bbet}-w_0^{\bbet}\,\1)
\end{array}\\ 
t_{\soul{A}}\in\cg_+ &\rightarrow&
\vec{T}_R(t_{\soul{A}})=\frac{i}{2} \left({f_{\soul{A}A}}^j{f_{jB}}^C
y_C\prt^A\prt^B  +({f_{\soul{A}A}}^j{f_{jB}}^B+
{f_{\soul{A}B}}^j{f_{jA}}^B)\prt^A\right).\, \1+\\
&&\phantom{\vec{T}_R(t_{\soul{A}})=} 
+R_{{\soul{A}}\bbet}(y_A,\prt^B).(M^{\bbet}-w_0^{\bbet}\,\1) 
\enano
where $M^{I}$ is the $d\times d$ matrix representing $W^{I}$, \1\ is
the $d\times d$ identity matrix, and $R_{\bal I}(y_A,\prt^B)$ are
polynomials in the differentials.
The above operators will act on vectors:
\[
\vec{\vph}(y_1,\dots,y_s)\ =\ \left(\begin{array}{c} 
\vph_1(y_1,\dots,y_s) \\ \vph_2(y_1,\dots,y_s) \\ 
\vdots \\ \vph_d(y_1,\dots,y_s) \end{array}\right)
\]

\subsection{Example: \su}
We continue the example of section \ref{Sectsu2}.
We denote an element of the group $SL(2)$ by its coordinates
\be
g(x^-,x^0,x^+)=e^{x^-t_-}e^{x^0t_0}e^{x^+t_+}
\ee
so that the functions read $\vph(x^-,x^0,x^+)$ on the regular
representation. In the differential representation, we have functions
$\wt{\vph}(y_-)\equiv\wt{\vph}(y)$ and we have
\be
T_0(t_-)=-iy \mb{;} T_0(t_0)=-y\prt -1
\mb{and} T_0(t_+)=-iy\prt^2 -2i\prt
\ee

We use the formula computed in section \ref{Sectsu2} to construct the
\cw-realization for \su. We have seen that, at the quantum level, 
$w=t_0^2+t_0+t_-t_+$
is the generator of the commutant of $\{t_0,t_-\}$ (see equation
(\ref{eq:1.5})).  We start with a
representation $T_0$ corresponding to a value $w_0$
of the Casimir operator. Then, we construct a new representation $T$
keeping for $t_0$ and $t_-$
the form $T_0$, $T(t_0)=T_0(t_0)$ and $T(t_-)=T_0(t_-)$, and
  inverting the formula (\ref{eq:1.5}) for $t_+$:
\beano
w-w_0 &=& \left[T(t_0)^2+T(t_0)+T(t_-)T(t_+)\right]-
\left[T_0(t_0)^2+T_0(t_0)+T_0(t_-)T_0(t_+)\right] \nonu
&=& T_0(t_-)\left[T(t_+)-T_0(t_+)\right] \nonu
&\Rightarrow& T(t_+)=T_0(t_+)+\frac{w-w_0}{T_0(t_-)}
\enano
Using the differential representation, we get
$w_0=0$ and
\be
T(t_+)=-iy\prt^2 -2i\prt+i\frac{w}{y} \mb{with} w\in\R  \label{toto}
\ee
Thus (\ref{toto}), together with $T(t_-)=-iy$ and $T(t_0)=-y\prt -1$ is a
representation of \su\ with a value $w$ for the Casimir operator.
Note that, for \su, a similar (up to a change of variable $y\leadsto x^2$)
realization has been given in \cite{LM}.

\section{Application to Poincar\'e and $so(4,2)$ algebras\label{sect3}}
We now apply the results of the previous sections to 
get realizations of the conformal
$so(4,2)$ algebra (and of its Poincar\'e subalgebra). We will show that
the \cw-realization of this latter provides the
Pauli-Lubanski-Wigner quadrivector $W^\mu$ in a very natural way. A
generalization to the whole $so(4,2)$ algebra will exhibit two new 
quadrivectors which,
together with $P^\mu$ and $W^\mu$, form a basis of the Minkowski space.
The scalar products of these vectors give all the physical quantities
such as the mass or the spin.

\subsection{Description}
The $so(4,2)$ algebra is known as 
the conformal algebra in four dimensions, in the
Minkowski  space, \ie with the metric
$g_{\mu\nu}=diag(1,-1,-1,-1)$. It is a non-compact form of
$so(6)$, with maximal compact subalgebra $so(4) \oplus so(2)$.
 Note 
that $s\ell(4)$ is also a non-compact version of $so(6)$, but with
maximal compact  subalgebra $so(4)$. Thus, one can connect $s\ell(4)$ to 
$so(4,2)$ through a
compactification process, a property which is useful for calculations.

The algebra $so(4,2)$ has fifteen generators:
\be
\begin{array}{ll}
\,\mbox{ Four translations: $P_{\mu}$ with $\mu=0,1,2,3$ which form }
& \cg_-\\ 
\left.\begin{array}{l}
\mbox{Three rotations and three boosts: $M_{ij}$ and $M_{0i}$ respectively
       ($i,j=1,2,3$)}\\
\mbox{One dilatation: $D$}
 \end{array}\right\} & \cg_0 \\
\,\mbox{ Four special conformal transformations: $K_{\mu}$ with $\mu=0,1,2,3$
which form } & \cg_+
\end{array}
\ee
Rotations and boosts constitute the Lorentz algebra $\cg_0=so(3,1)$, and 
together with the translations those generators form the \pcr
algebra.
In the following we will denote by 
$\cp=\cg_0\oplus\cg_{+}$ the corresponding parabolic algebra. This
latter  decomposition is actually induced by a natural gradation of 
the algebra 
with respect to the element $-iD$, which
assigns a grade $-1$ to $P_\mu$, $0$ to $M_{\mu\nu}$, and $+1$ to
 $K_\mu$. The commutation relations read:
\be
\begin{array}{l}
{[P_\mu, P_\nu]}=0 \mb{ } {[M_{\mu\nu}, P_\rho]}= i(g_{\nu\rho}P_\mu- 
g_{\mu\rho}P_\nu) \mb{ } {[D,P_\mu]}=-iP_\mu \\
{[M_{\mu\nu}, M_{\rho\sig}]}= i(g_{\nu\rho}M_{\mu\sig}-
g_{\mu\rho}M_{\nu\sig}+g_{\mu\sig}M_{\nu\rho}-g_{\nu\sig}M_{\mu\rho})
\mb{ } {[D, M_{\mu\nu}]}=0 \\
{[K_\mu, K_\nu]}=0 \mb{ } {[M_{\mu\nu}, K_\rho]}= i(g_{\nu\rho}K_\mu- 
g_{\mu\rho}K_\nu) \mb{ } {[D,K_\mu]}=iK_\mu \\
{[P_\mu,K_\nu]}=2i(-M_{\mu\nu}+g_{\mu\nu}D) 
\end{array}
\ee
$so(4,2)$ possesses three Casimir operators of degree 2, 3 and 4. We will call
them $C_2$, $C_3$ and $C_4$ respectively. 

We will need a frame $(e_0,e_1,e_2,e_3)$ for the Minkowski space:
\be
\begin{array}{l}
e_0=(e_0)_\mu=(1,\vec{0}) \\
e_j=(e_j)_\mu=(0,\vec{e_j})\ \ j=1,2,3 \end{array}
\mb{with} \left\{
\begin{array}{l}
\vec{e_1}=(1,0,0) \\ \vec{e_2}=(0,1,0) \\ \vec{e_3}=(0,0,1)
\end{array} \right.
\ee
which satisfies
\be
e_\mu\cdot e_\nu=(e_\mu)^\rho (e_\nu)^\sig g_{\rho\sig}=g_{\mu\nu}
\mb{and}\ (e_\mu)^\rho (e_\nu)^\sig g^{\mu\nu}=g^{\rho\sig}
\ee

\subsection{Momentum representation}
We first construct the differential representation associated to $\cg_-$
(see section \ref{diffReal}).
We start with a coordinate representation, where functions depend on
${x}=(x^\mu)$ and perform a Fourier transformation: 
$$\wt{\vph}( p) = \int d^s x\ e^{i p \cdot x}\ \soul{\vph}(x)$$ 
The coordinate differential representation on the space $G/P$:
\be
\begin{array}{rclrcl}
P_{\mu} &=& {\displaystyle i \frac{\partial}{\prt x^{\mu}}} & \quad
M_{\mu\nu} &=& i {\displaystyle (x_{\mu} \frac{\prt}{\prt x^{\nu}} - x_{\nu}
\frac{\prt}{\prt x^{\mu}})} \\ [.3cm]
D &=& {\displaystyle i x \cdot \frac{\prt}{\prt x}} &
K_{\mu} &=& {\displaystyle i(2 x_{\mu}\, x \cdot \frac{\prt}{\prt x} -  x^{2}
\frac{\prt}{\prt x^{\mu}}) }
\end{array} 
\ee
is thus converted into the momentum representation: 
\be
\begin{array}{rclrcl}
P_{\mu} &=& p_{\mu} &
M_{\mu\nu} &= & i (p_{\mu}\prt_{\nu} - p_{\nu}\prt_{\mu} ) \\ [.3cm]
D &=& -i ( p\cdot\prt +4) &
K_{\mu} &=&  (p_{\mu}\Box -2 p\cdot\prt\prt_{\mu}  - 8\prt_\mu ) \label{diff3}
\end{array} 
\ee
where $\prt_\mu$ stands for $\frac{\prt}{\prt p^{\mu}}$, $p$ is the
quadrivector $(p_\mu)$, $\prt$ the quadrivector $(\prt_\mu)$, and
$\Box=\prt\cdot\prt=g^{\mu\nu}\prt_\mu\prt_\nu=\prt^\mu\prt_\mu$. 

\subsection{The \cw-algebra \label{sect4:3} }
We wish to construct the commutant of the $P_\mu$ in order to build a
\cw-representation for $\cg=so(4,2)$.
We have seen that $D$ is the right grading operator, since
$\cg_-=\{P_\mu\}$.
In the case we consider $s\ell(4)$ the maximally non-compact form of $\cg$,
this gradation corresponds to the model $\cw(s\ell(4),2\su)$. Then referring
to the formula (\ref{gtild}) we can take $\tcg=\cg_-\oplus\tch$ where
\be
\tch=\{M_{13}-M_{01},\ M_{23}-M_{02},\ M_{03},\ D\}.
\ee

The commutant of \tcg\ can be seen as a compactified form\footnote{Let us
mention the recent work \cite{MadsEv} on the real forms of \cw-algebras.}
of the algebra $\cw(s\ell(4),2\su)$. Indeed, calculations can be done
in 
an easier way
when using the $s\ell(4)$ algebra (instead of $so(4,2)$) and then
compactifying the resulting algebra. This algebra is a polynomial 
deformation of $\cg_0=so(3,1)\oplus so(1,1)$. It contains seven generators:
three generators $J_k$, $k=1,2,3$ of degree one in the $so(4,2)$ elements, 
which form $so(3)$, a Lie 
subalgebra of the \cw-algebra, and also four generators $C_2$ and $S_k$,
$k=1,2,3$ of degree two. $C_2$ is the second order Casimir of 
$so(4,2)$:
\be
C_2 = \frac{1}{2}(g^{\mu \rho}g^{\nu \sig}M_{\mu \nu} M_{\rho \sig}
+g^{\mu \nu}(P_{\mu}K_{\nu}+K_{\mu}P_{\nu}))-D^2
\ee
For the clarity of the presentation, we postpone the
expressions of the other generators in terms of the $so(4,2)$ generators
(formulae (\ref{defJ}), (\ref{defSig}) and (\ref{defS})).  

The commutation relations of the \cw-algebra read:
\bea
{[J_j,J_k]} &=&i \epsilon_{jkl}J_l \nonu
{[J_j,S_k]} &=&i \epsilon_{jkl}S_l \nonu
{[S_j,S_k]} &=&-i \epsilon_{jkl} \left( 2 (J_1^2+J_2^2+J_3^2) -C_2-4 \right)J_l
\ena

\subsection{The \cw-realization}
Now using technics summarized in section \ref{sect2:3} as well as the results 
of \ref{sect4:3}, one can determine the general expressions of the conformal
algebra generators. The generators corresponding to $\tcg$ are unchanged with
respect to the differential realization (\ref{diff3}).

One remarks that the $so(4,2)$ generators are indeed linear in the $J_i$ 
$(i=1,2,3)$ and $Z_k$ $(k=0,1,2,3)$ -see (\ref{basis})- which also form a basis
of the \cw-algebra.

One notes also that only the $\vec{J}$-part of the \cw-algebra shows up in the
Poincar\'e part: this will be discussed in the next paragraph.

One can check that every \cw-generator reduces to zero in the differential
realization (\ref{diff3}).

Finally let us emphasize that the realization given just below corresponds to
values of $p^2>0$; therefore the quantities $p_0+p_3$ and $p^2$ appearing in
the denominators never vanish.
Let us now give the realization that one gets from our "\cw-approach"
(\1\ denotes the identity matrix):

\bea
P_\mu &=& p_\mu\, \1 \mb{ }\mu=0,1,2,3 \label{Pmu}  \\
M_{12} &=& i (p_{1}\prt_{2} - p_{2}\prt_{1} )\, \1 +J_3 \label{M12}  \\
M_{13}&=& i (p_{1}\prt_{3} - p_{3}\prt_{1} ) \, \1
-\frac{\sqrt{p^2}}{p_0+p_3}J_2 -\frac{p_2}{p_0+p_3}J_3 \\
M_{23} &=& i (p_{2}\prt_{3} - p_{3}\prt_{2} )\, \1
+\frac{\sqrt{p^2}}{p_0+p_3}J_1 +\frac{p_1}{p_0+p_3}J_3 \\
M_{01} &=& i (p_{0}\prt_{1} - p_{1}\prt_{0} ) \, \1
-\frac{\sqrt{p^2}}{p_0+p_3}J_2 -\frac{p_2}{p_0+p_3}J_3 \\
M_{02} &=& i (p_{0}\prt_{2} - p_{2}\prt_{0} )\, \1
+\frac{\sqrt{p^2}}{p_0+p_3}J_1 +\frac{p_1}{p_0+p_3}J_3 \\
M_{03} &=& i (p_{0}\prt_{3} - p_{3}\prt_{0} ) \, \1  \label{M03} \\
D &=& -i ( p\cdot\prt +4)\, \1 \label{dilat}   \\
K_0 &=& (p_{0}\Box -2 p\cdot\prt\prt_{0}  - 8\prt_0 )\, \1
-\frac{2}{p_0+p_3}Z_3 +\frac{p_0}{p^2}Z_0 
+\frac{1}{(p_0+p_3)\sqrt{p^2}}(p_1 Z_1 +p_2 Z_2) +\nonu
&& -\frac{2i \sqrt{p^2}}{p_0+p_3} 
\left( -(\frac{5}{2} \frac{p_2}{p^2} +\prt_2 )J_1 
+(\frac{5}{2} \frac{p_1}{p^2} +\prt_1)J_2 \right) 
- \frac{2i}{p_0+p_3}(p_2\prt_1-p_1\prt_2)J_3 \label{K0}   \\
K_1 &=& (p_{1}\Box -2 p\cdot\prt\prt_{1}  - 8\prt_1 )\, \1
+\frac{1}{\sqrt{p^2}}Z_1 +\frac{p_1}{p^2}Z_0 +\nonu
&& -\frac{2i \sqrt{p^2}}{p_0+p_3} 
\left( \frac{5}{2} \frac{p_0+p_3}{p^2} +\prt_0 +\prt_3 \right) J_2
-2i \left( \frac{p_2}{p_0+p_3}(\prt_0+\prt_3)-\prt_2 \right)J_3 \\
K_2 &=& (p_{2}\Box -2 p\cdot\prt\prt_{2}  - 8\prt_2 )\, \1
+\frac{1}{\sqrt{p^2}}Z_2 +\frac{p_2}{p^2}Z_0 +\nonu
&& +\frac{2i \sqrt{p^2}}{p_0+p_3} 
\left( \frac{5}{2} \frac{p_0+p_3}{p^2} +\prt_0 +\prt_3 \right) J_1
+2i \left( \frac{p_1}{p_0+p_3}(\prt_0+\prt_3)-\prt_1 \right)J_3 \\
K_3 &=& (p_{3}\Box -2 p\cdot\prt\prt_{3}  - 8\prt_3 )\, \1
+\frac{2}{p_0+p_3}Z_3 +\frac{p_3}{p^2}Z_0
-\frac{1}{(p_0+p_3)\sqrt{p^2}}(p_1 Z_1 +p_2 Z_2) +\nonu
&& +\frac{2i \sqrt{p^2}}{p_0+p_3} 
\left( -(\frac{5}{2} \frac{p_2}{p^2} +\prt_2 )J_1 
+(\frac{5}{2} \frac{p_1}{p^2} +\prt_1)J_2 \right) 
+ \frac{2i}{p_0+p_3}(p_2\prt_1-p_1\prt_2)J_3 \label{K3}
\ena
where 
\be
\begin{array}{ll}
Z_1=2S_1+J_3J_1+J_1J_3  &  Z_2=2S_2+J_3J_2+J_2J_3  \\
Z_3=S_3-(J_1^2+J_2^2)  & Z_0=2S_3+C_2-J_3^2-2(J_1^2+J_2^2)
\label{basis}
\end{array}
\ee

\subsection{The Poincar\'e algebra\label{poinc}}
Let us focus on the expressions of the Poincar\'e generators (\ref{Pmu})-
(\ref{M03}). We recall the expressions of the Pauli-Lubanski-Wigner 
quadri-vector
$W^\mu=\half\eps^{\mu\nu\rho\sig} P_\nu M_{\rho\sig}$ which satisfies:
\be
{[W_\mu, P_\nu]}=0 \mb{ } 
{[W_\mu, W_\nu]}=i\eps_{\mu\nu\rho\sig} W^\rho P^\sig \mb{ }
 W\cdot P=0
\ee
It is well-known that the irreducible representations of the \pcr algebra are
labelled by the eigenvalues of $P^2=p^2$ and $W^2=-s(s+1)p^2$, 
where $s$ is the spin of the particle. Because of the relation 
$W\cdot P=0$, the quadrivector $W^\mu$ possesses only 3 independant
components. These 3 independant components generate the spin algebra
$so(3)$ when $p^2$ is positive \cite{MoSt}.
This is recovered in a very natural way in our \cw-algebra framework.
Indeed, the generators $J_k$ really play the role of the spin generators, since
they can be rewritten as:
\be
J_k=-\frac{1}{m} n_k\cdot W=- \frac{1}{m} (n_k)^\mu W_\mu
\mb{and} W^\mu=-m\sum_{k=1}^{k=3} (n_k)^\mu J_k 
\mb{(since $P\cdot W=0$)} \label{defJ}
\ee
\be
\mbox{with} \qquad m=\sqrt{p^2} \qquad \mbox{and} \qquad 
W \cdot W = -P^2 \vec{J}^2 \label{WW}
\ee
and where we have introduced
 the frame of the "particle" of momentum $p$:
\bea
n_0 &=& (n_0)_{\mu} = \frac{1}{m}\, p\ =\ 
(\frac{p_0}{m},\frac{p_1}{m},\frac{p_2}{m},\frac{p_3}{m}) 
\label{eq:no}\\
n_1 &=& (n_1)_{\mu} = \frac{1}{p_0+p_3}(p_1,p_0+p_3,0,-p_1) 
\label{eq:n1}\\
n_2 &=& (n_2)_{\mu} = \frac{1}{p_0+p_3}(p_2,0,p_0+p_3,-p_2) 
\label{eq:n2}\\
n_3 &=& (n_3)_{\mu} = \frac{-m}{p_0+p_3}(1,0,0,-1) +\frac{1}{m}\, p
\label{eq:n3}
\ena
which obeys to 
\be
n_\mu\cdot n_\nu=(n_\mu)^\rho (n_\nu)^\sig
g_{\rho\sig}=g_{\mu\nu}  
\ee
and also to
\be
(n_\mu)^\rho (n_\nu)^\sig g^{\mu\nu}=
g^{\rho\sig}  
\ee
The Lorentz transformation, denoted by $[p]$ in \cite{MoSt} and $L(p)$ in
\cite{Mack}, which moves the rigid referential frame $(e_0,e_1,e_2,e_3)$ to 
the $p$-frame $(n_0,n_1,n_2,n_3)$:
\be
\begin{array}{l}
\displaystyle (e_0,e_1,e_2,e_3)\ \stackrel{L(p)}{\longrightarrow}\
(n_0,n_1,n_2,n_3)\\ 
\end{array} \label{defA}
\ee
can explicitely be written as a rotation followed by a boost
\be
\begin{array}{l}
\displaystyle L(p)=B(p)R(p) \mb{with} B(p)=exp(-i\lda\
\frac{\vec{p}\cdot\vec{B}}{\|\vec{p}\|}) 
\mb{and} R(p)=exp(-i\theta\
\frac{\vec{p}_\perp\cdot\vec{R}}{\|\vec{p}_\perp\|})\\
\displaystyle \mb{with} sh(\lda)=\frac{\|\vec{p}\|}{m},\
cos(\theta)=1-\frac{\|\vec{p}_\perp\|^2}{(m+p_0)(p_0+p_3)}
\mb{and} \vec{p}_\perp=(p_2,-p_1,0)
\end{array} \label{defU}
\ee
where we have defined $\vec{B}=(M_{01},M_{02},M_{03})$ and 
$\vec{R}=(M_{23},M_{31},M_{12})$.
This transformation also relates the three vector $(J_i)$ with the four vector
$(W_{\mu})$:
\be
mJ=(0,m\vec{J})\ \stackrel{L(p)}{\longrightarrow}\ W=(W_\mu)
\ee

Actually, $L(p)$ belongs to the set of Lorentz transformations which send the 
four vector $(m,0,0,0)$ into $p=(p_0,p_1,p_2,p_3)$. It is through $L(p)$
that the representations of the Poincar\'e group can be constructed from 
representations of the rotation subgroup. Indeed, following \cite{MoSt} and
\cite{Mack}, the Lorentz transformation $\Lambda$ acting on function 
$\tilde{\varphi}$
of the $p$-variable in the $U$ representation:
$$^{\Lambda}\tilde{\varphi}(p)=U(\Lambda)\tilde{\varphi}(\Lambda^{-1}p)$$
writes more conveniently on the Wigner functions $\psi$ defined by 
$$\psi(p)=U(L(p)^{-1})\tilde{\varphi}(p)$$
as
$$ \left( ^{\Lambda}\psi \right) (p)=U\left( L(p)^{-1}\Lambda L(\Lambda^{-1}p)
\right) \psi(\Lambda^{-1}p).$$
We recognize in the product $L(p)^{-1}\Lambda L(\Lambda^{-1}p)$ a Wigner 
rotation, element of the $(m,0,0,0)$-vector stabilizer, itself isomorphic to 
the $SO(3)$ group when $p^2>0$.

Let us emphasize that the transformation $L(p)$ can be chosen up to a 
$R$-rotation leaving $(m,0,0,0)$ invariant. In particular, the $L(p)$ used in
the computations of \cite{Mack} is the pure boost $B(p)$. This of course 
produces a new
but equivalent frame $(n_0',n_1',n_2',n_3')$ and different expressions for the
Lie algebra realization.

\subsection{Generalization to $so(4,2)$\label{reste}}

We now look at the other generators of the conformal algebra. We proceed
as for the \pcr algebra. In the same way we have introduced the 
Pauli-Lubanski-Wigner vector $W^\mu$, let us define:
\be
\Sigma_\mu=-W^2\, P_\mu +
P^2\left[ P^\A M_{\A\mu}(D+i)- \half(P_\mu\, P\!\cdot\! K-
P^2\, K_\mu)-\half\eps_{\mu\nu\rho\sig} W^\nu M^{\rho\sig} \right]
\label{defSig}
\ee
It satisfies
\be
\begin{array}{ll}
{[\Sigma_\mu, P_\nu]}=0 & \Sigma\cdot P=0 \\
{[\Sigma_\mu, W_\nu]}=i\eps_{\mu\nu\rho\sig} \Sigma^\rho P^\sig & \\
{[\Sigma_\mu, \Sigma_\nu]}=i\eps_{\mu\nu\rho\sig}\, P^2\, W^\rho P^\sig 
(P^2 (4+C_2)+ 2 W^2)\ \ & 
\end{array}
\ee
The generators $S_i$ are connected to the quadrivector $(\Sigma^\mu)$ through
\be
S_k=-\frac{1}{m^3}n_k\cdot \Sigma=-\frac{1}{m^3} (n_k)^\mu \Sigma_\mu
\mb{and} \Sigma^\mu=-m^3\,\sum_{k=1}^{k=3} (n_k)^\mu S_k 
\mb{(since $P\cdot \Sigma=0$)} \label{defS}
\ee
and also:
\be
m^3S=(0,m^3\vec{S})\ \stackrel{L(p)}{\longrightarrow}\
\Sigma=(\Sigma_\mu) \ee
Let us remark that the third and fourth Casimir operators of $so(4,2)$
can be expressed as $\cw$-polynomials: 
\begin{eqnarray}
\displaystyle \vec{J} \cdot \vec{S} &=& -\frac{W \cdot \Sigma}{(P^2)^2}
=-\frac{1}{2}C_3 \label{JS} \\
\displaystyle \vec{S}^2 +\vec{J}^2(C_2-\vec{J}^2+2) &=& \frac{1}{(P^2)^3}\left(
\Sigma^2+P^2 W^2 \left( W^2+P^2(C_2+2) \right)\right)=-\frac{1}{4}C_4+
\frac{2}{3}C_2  \nonumber
\end{eqnarray}
with the following expression for the Casimir operators:
\bea
C_3 &=& -\frac{1}{4} DM_{\mu\nu}\wt{M}^{\mu\nu}+ \half(K\cdot W+P\cdot W) \\
C_4 &=& 2(W\cdot V+V\cdot W)+\half(P^2 K^2+K^2 P^2) -(P\cdot K)^2 
+\frac{1}{16} 
(M_{\mu\nu}\wt{M}^{\mu\nu})^2 +\nonu
&&-2(D^2+2) M_{\mu\nu}{M}^{\mu\nu}+4D P^\mu K^\nu M_{\mu\nu} -8i D
(C_2-D^2)+\frac{4}{3} C_2 -20 D^2
\ena
where we have introduced
\be
V^\mu = \half\eps^{\mu\nu\rho\sig}K_\nu M_{\rho\sig} \mb{and}
\wt{M}^{\mu\nu} = \eps^{\mu\nu\rho\sig}M_{\rho\sig}
\ee

Note that $(e_0, J, S)$ on the one hand, and $(P, W, \Sigma)$
on the other hand form three independant quadrivectors: it is natural to
look for a fourth quadrivector $X=(0,\vec{X})$ such that $(e_0, J, S,
X)$ is a frame. Then, we will get a quadrivector $\chi$ such that 
\be
m^4 X \stackrel{L(p)}{\longrightarrow} \chi \mb{and} (P, W, \Sigma, \chi)
\mbox{ frame}
\ee
Calculations show that
\be
\begin{array}{l}
X_1=(J_2S_3+S_3J_2)-(J_3S_2+S_2J_3)-3i S_1 \\
X_2=(J_3S_1+S_1J_3)-(J_1S_3+S_3J_1)-3i S_2 \\
X_3=(J_1S_2+S_2J_1)-(J_2S_1+S_1J_2)-3i S_3 
\end{array} \mb{with} \vec{X}=(X_1,X_2,X_3)
\ee
is such a vector. Note that the leading term of $\vec{X}$ is a
non-commutative version of the exterior product $\vec{J}\wedge\vec{S}$.
From equations (\ref{WW}) and (\ref{JS}), one sees that all the "physical 
informations" of the \pcr algebra as
well as of the conformal algebra $so(4,2)$ are contained in these frames.

\subsection{Case $m=0$}
Up to now, the results we have presented are valid for $p^2>0$.
One can wonder whether there exists some limiting process that
could lead to the case of zero mass representations. At that point, 
we have to make a distinction between the $so(4,2)$ algebra
and its \pcr subalgebra. Indeed, in the \pcr algebra, the mass is an
invariant ($p^2=m^2$), so that it is very natural to look at the limit
point $m=0$. In the $so(4,2)$ algebra, this is no more the case, but
the  sign $\eps(p^2)$ is still an invariant. By 
sign of $p^2$, we mean $\eps(p^2)=+,0,-$ when $p^2$ is respectively
positive, zero or negative. One easily guesses that a limiting
process on the three points $\eps=+1,0,-1$ is hard to find. In other
words, in the \pcr algebra, the (positive mass) irreducible representations are
hyperboloids in the Minkowski space, while in the $so(4,2)$ algebra, 
the irreducible
representation $\eps=+1$ is the whole interior of the light cone. Then, 
if, in the \pcr algebra, it is natural to look at the light cone as the
limit of the family of hyperboloids when $m\rightarrow0$, 
 we have to face a problem of dimensional reduction  of the
space of representations in the $so(4,2)$ (see \cite{MT} for a direct
treatment of the case $m=0$ in $so(4,2)$).

Thus, we will focus below on the \pcr algebra to study the limit
$m\rightarrow 0$. The \cw-algebra then reduces to a three-dimensional
algebra since the generators $\vec{S}$ and $C_2$ do not enter in the
realization (\ref{Pmu}-\ref{M03}).
Moreover,
in the case $m=0$, one cannot introduce the vector $\vec{J}$ and the
frame attached to the particle, which is now of zero mass. 
However, we can define:
\be
Q_j=n_j\cdot W \mb{for} j=1,2 \mb{while} 
Q_3=\frac{1}{p_0+p_3}(e_0-e_3)\cdot W  \label{defQ}
\ee
which form an $E(2)$ subalgebra
\be
[Q_3, Q_1]=iQ_2 \mb{} [Q_3, Q_2]=-iQ_1 \mb{} [Q_1, Q_2]=0
\ee   
 in accordance with
the general results on zero mass particles. Moreover, (\ref{defQ})
shows that for $W^\mu$ proportional to $P^\mu$, and using the 
definitions (\ref{eq:n1}-\ref{eq:n2}),
 we have $Q_1=Q_2=0$. Let us remind that this latter condition is the
 one we need to select the physical cases when $m=0$, \ie the cases of
 finite spin (more exactly helicity). By definition, these latter cases
 are  the  ones  that correspond to finite dimensional representations of
 the $E(2)$ algebra. Thus, they are also natural in  the
 \cw-realizations, since we consider a priori matricial representations
 of  the \cw-algebra. In that case, the realization reads:
\bea
P_\mu &=& p_\mu\, \1 \mb{ }\mu=0,1,2,3   \\
M_{12} &=& i (p_{1}\prt_{2} - p_{2}\prt_{1} )\, \1 +Q_3   \\
M_{13}&=& i (p_{1}\prt_{3} - p_{3}\prt_{1} ) \, \1
 -\frac{p_2}{p_0+p_3}Q_3 \\
M_{23} &=& i (p_{2}\prt_{3} - p_{3}\prt_{2} )\, \1
 +\frac{p_1}{p_0+p_3}Q_3 \\
M_{01} &=& i (p_{0}\prt_{1} - p_{1}\prt_{0} ) \, \1
 -\frac{p_2}{p_0+p_3}Q_3 \\
M_{02} &=& i (p_{0}\prt_{2} - p_{2}\prt_{0} )\, \1
+\frac{p_1}{p_0+p_3}Q_3 \\
M_{03} &=& i (p_{0}\prt_{3} - p_{3}\prt_{0} ) \, \1   
\ena
where $Q_3$ corresponds to the helicity. Let us note that although the
denominator $p_0+p_3$ may vanish when $m=0$, the above expressions 
can still be well-defined using limits $p_0+p_3\rightarrow0$ and
$p_j=o(p_0+p_3)$ ($j=1,2$).

\subsection{Comparison with induced representations}
It is time to compare the above obtained $\cw$-realization of the $so(4,2)$ 
algebra with the usual ones constructed via the induced representation method.
As could be expected, the results are equivalent. More is given in \cite{Mack}
where the classification of all the unitary ray representation of the $SU(2,2)$
group with positive energy is achieved. It will then be possible for us, using
these results, to select the finite dimensional representations of the
$\cw$-algebra which lead to such unitary representations of the conformal
one, and therefore to unitary representations of the conformal group.

Let us start by introducing a realization of the $\cw$-algebra under 
consideration known as the Miura realization, and particularly useful in
what follows. Indeed, it is the Miura realization of the $\cal W$-algebra
which can be unearthed in \cite{Mack}. Its knowledge also provides a simple
algorithm for the construction of representations of the $\cw$-algebra. 

\subsubsection{The Miura realization}
The Miura transformation provides a realization of the $\cw$-algebra relative
to the gradation $\cg=\cg_{-1}\oplus \cg_{0}\oplus \cg_{+1}$ in terms of 
elements of the enveloping algebra $\cu(\cg_0)$ \cite{BT}. There, the
 $\cg_0$ part is the Lorentz algebra to which have been added the dilatation:
$so(3,1) \oplus so(1,1)$, and the Miura realization reads simply:
\be
\vec{J}=\vec{R} \qquad  \vec{S}=\vec{R} \times \vec{B} -i(D-1)\vec{B}
\qquad C_2=\vec{R}^2-\vec{B}^2+D(D-4). \label{Miura}
\ee
using again the notations of section \ref{poinc} for the Lorentz generators. 
One immediately remarks that this realization is much simpler than the one 
above determined and which involves all the $so(4,2)$ generators: (\ref{defJ})
and (\ref{defS}). Thus, to each finite dimensional $\cg_0$ representation
can be associated, via equation (\ref{Miura}), a finite dimensional $\cw$
representation. Labelling in a rather natural way $(j_1,j_2;d)$ an
$so(3,1) \oplus so(1,1)$ irreducible representation, the center element 
$C_2$ in $\cw$
will get the scalar value:
$$c_2=2j_1(j_1+1)+2j_2(j_2+1)+d(d-4).$$ 

\subsubsection{Unitary representations}
The unitary representations with positive energy of the universal covering 
group $\hat{G}$ of $G=SU(2,2)$ are presented in \cite{Mack} -see also
\cite{Ruhl}. We note the chain of isomorphisms:
\be
\left(
\begin{array}{c} \mbox{conformal group of} \\ \mbox{Minkowski space} 
\end{array} \right)
\simeq SO(4,2)/ \, \Z_2 \simeq SU(2,2)/ \, \Z_4 \simeq \hat{G}/
(\Gamma_1  \times \Gamma_2)
\ee
where $\Gamma_1 \simeq \Z_2$ and $\Gamma_2 \simeq \Z$.
Such unitary representations possess a lowest weight, and are interpolation
of the holomorphic discrete series. These latter are
realized by considering functions on the bounded symmetric domain 
$SU(2,2)/K$, where $K=SU(2) \times SU(2) \times U(1)$ is the maximal compact 
subgroup of $SU(2,2)$. Actually they are induced by the finite
dimensional (unitary) representations of $K$.

However, Mack \cite{Mack} described them as induced representations by the 
parabolic 
subgroup $P=\Gamma_2 MAN$, where following his notations $M$, $A$, $N$ are
the subgroups of Lorentz, dilatation and special conformal transformations
respectively. The Lie algebra of $P$ is exactly $\cg_0 \oplus \cg_+$ in our
language. We note that $P$ is actually the little group of 
the point $x=0$ in the compactified Minkowski space, itself isomorphic to 
$\hat{G}/P$. The induced representations on this
space are then constructed in the usual way. We take finite dimensional
representations $D^{\lambda}$ of the parabolic subgroup $P$, where $N$ acts 
trivially.
They are labelled by two non-negative (half-)integers $(j_1,j_2)$, associated 
to spinor representations of the Lorentz group $D^{j_2j_1}$, and by $d$ a 
real number associated to the dilatation:
$$D^{\lambda}(man)=|a|^{d-2}D^{j_2j_1}(m).$$
It acts on a $(2j_1+1)(2j_2+1)$
dimensional vector space $E^{\lambda}$. From now on we will disgard the action 
of the discrete central part $\Gamma$, which of course will not affect the Lie
algebra realization.
 We then consider the space of functions
$\varphi$ on the group $G$ with values in $E^{\lambda}$ which have the 
covariance property 
$$\varphi(gman)=D^{\lambda}(man)^{-1} \varphi(g) \quad \forall g \in SU(2,2).$$
The conformal group acts on these functions by the left regular representation:
$$(T(g)\varphi)(g')=\varphi(g^{-1}g').$$
Since the decomposition $g=g_-man$ is unique, functions $\varphi$ can be 
viewed as functions $\varphi(x)$ defined on $G/P$, the Minkowski space. The 
action of $SU(2,2)$ then takes the form:
$$(T(g)\varphi)(x)=|a|^2 D^{\lambda}(man)^{-1}\varphi(x')$$ 
where $x' (\in G_-),m,a,n$ are defined through the unique decomposition 
$g^{-1}x=x'man$. 
Note that this way to describe unitary representations of $SU(2,2)$ is unusual
since we induced by non-unitary representations of a parabolic subgroup which
is not cuspidal \cite{Knapp}. By cuspidal, we mean that the Cartan subalgebra
of the semi-simple part $M$ of $P$ is made only with compact generators, which
is not the case here.

To get the action of the Lie algebra $so(4,2)$, one has simply to take the 
infinitesimal form:
\be
\begin{array}{rclrcl}
P_{\mu} &=& {\displaystyle i\frac{\partial}{\partial x^{\mu}}} & 
M_{\mu \nu} &=& {\displaystyle i(x_{\mu} \frac{\partial}{\partial x^{\nu}} 
-x_{\nu} \frac{\partial}{\partial x^{\mu}} -i \Sigma_{\mu \nu}) } \\ [.3cm]
D &=& {\displaystyle i(4-d+x \cdot \frac{\partial}{\partial x}) } & \quad
K_{\mu} &=& {\displaystyle i((8-2d)x_{\mu}+2x_{\mu}x \cdot 
\frac{\partial}{\partial x} 
-x^2 \frac{\partial}{\partial x^{\mu}} -2ix^{\nu} \Sigma_{\mu \nu}) }
\end{array} 
\ee
Here $\Sigma_{\mu \nu}$ is a $(2j_1+1)(2j_2+1)$ representation of the Lorentz
algebra with
$R_i=\frac{1}{2}\epsilon_{ijk} \Sigma^{jk}$ and $B_i=\Sigma_{0i}$.

The comparison with the $\cal W$-realization proceeds in several steps.
First of all, we have to perform a Fourier transformation in order to get
momentum representations with functions $\tilde{\varphi}(p)$:
\be
\begin{array}{rclrcl}
P_{\mu} &=& p_{\mu} & \quad 
M_{\mu \nu} &=& i(p_{\mu} \partial_{\nu} -p_{\nu} \partial_{\mu} 
-i\Sigma_{\mu \nu}) \\ [.3cm]
D &=& -i(p \cdot \partial +d) & 
K_{\mu} &=& p_{\mu} \Box -2p \cdot \partial \partial_{\mu} -2d \partial_{\mu}
-2i\Sigma_{\mu \nu} \partial^{\nu}
\end{array} 
\ee
Then we study the action of the conformal algebra on the functions
$\psi(p)$ defined by
$$\psi(p)= D^{j_2j_1}\!\left(L(p)^{-1}\right)\, m^{d-4}
\tilde{\varphi}(p)$$ where $L(p)$ is a Lorentz transformation that 
transports the frame 
$(e_0,e_1,e_2,e_3)$ to the frame $(n_0,n_1,n_2,n_3)$. 
If $X$ is the form of a generator of $so(4,2)$
acting on functions $\tilde{\varphi}(p)$ then, when acting on functions 
$\psi(p)$ it becomes
$$
\hat{X}= D^{j_2j_1}\!\left(L(p)^{-1}\right)\ m^{d-4}\, X \, m^{-(d-4)}\
D^{j_2j_1}\!\left(L(p)\rule{0mm}{1em}\right)
$$ 
Note that when restricted to the Poincar\'e algebra, since the mass is constant
on an irreducible representation, this whole trick amounts in working on the 
Wigner wave function (cf section \ref{poinc}). We then recover the well-known 
representations of 
the Poincar\'e group induced by the group of rotations. Explicitely, Lorentz 
transformations are given by
$$(T(\Lambda)\psi)(p) = D^{j_2j_1}\!\left(L(p)^{-1} \Lambda
L(\Lambda^{-1}p)) \psi(\Lambda^{-1}p\right)$$
where $D^{j_2j_1}(L(p)^{-1} \Lambda L(\Lambda^{-1}p))$ reduces simply to 
representations of spin $s=|j_1-j_2|,\dots,j_1+j_2$ of $so(3)$. This
realization of  the Poincar\'e 
algebra corresponds exactly to the form we find for the 
$\cal W$-representation (\ref{M12})-(\ref{M03}), by identifying $J_i$ 
with $R_i$.

For the other $so(4,2)$ generators, to calculate $\hat{X}$, we 
first compute $m^{d-4}Xm^{-(d-4)}$. Then,  
we perform a second order Taylor 
expansion of $L(p)$ around $(p_0,\vec{p})=(m,\vec{0})$. 
Finally, we get  
the general form of $\hat{X}$, using the fact that $K_{\mu}$ 
transforms as a four-vector under a Lorentz transformation, and $D$ as a
scalar.
We obtain for the dilatation the realization (\ref{dilat}) computed in
the \cw-representation.
We also get  the \cw-realization (\ref{K0})-(\ref{K3}) for 
$\hat{K_{\mu}}$, provided we identify 
\begin{equation}
\vec{J}=\vec{R} \qquad \vec{S}=\vec{R} \times \vec{B} -i(d-1)\vec{B}
\qquad C_2=\vec{R}^2-\vec{B}^2+d(d-4), 
\end{equation}
which is precisely the Miura realization given in (\ref{Miura}).
Thus, we have made the link between the $\cal W$-realization and the
induced representation technics.

As a further consequence, the unitarity conditions on $j_1$,$j_2$ and $d$ 
given in \cite{Mack} can be seen as conditions on the $\cw$-algebra 
representations. In particular the class of representations of positive
masses 
\be
\begin{array}{l}
d\geq j_1+j_2+2 \quad \mbox{with} \quad j_1j_2 \neq 0\\
d> j_1+j_2+1 \quad \mbox{with} \quad j_1j_2 = 0
\end{array}
\ee
gives rise respectively to the $C_2$ conditions:
\be
\begin{array}{l}
c_2 \geq 2j_1(j_1+1)+2j_2(j_2+1)+(j_1+j_2)^2-4\\
c_2> 3(j_1+j_2+1)(j_1+j_2-1)
\end{array}
\ee
For $j_1j_2 \neq 0$ and $d> j_1+j_2+2$, each 
 representation contains spins $s=|j_1-j_2|,\cdots,j_1+j_2$,
while for $d= j_1+j_2+2$ or $j_1j_2 = 0$ each representation contains
only the spin $s=j_1+j_2$.

\indent

\indent

\noindent { \Large\bf Acknowledgements}

\indent

We would like to thanks Philippe Caldero for valuable discussions about
commutants in Lie algebras.
We are also indebted to Raymond Stora for his interest and 
numerous comments on induced representations of the \pcr and conformal
algebras.

\end{document}